\documentclass{elsart}
\usepackage{graphicx}
\usepackage{amsmath}
\usepackage{amssymb}

\begin{document}

\begin{frontmatter}

\title{Measuring orbital interaction using quantum information theory}
\author[Marburg]{J\"org Rissler\corauthref{ich}},
\ead{rissler@staff.uni-marburg.de}
\author[Marburg]{Reinhard M. Noack},
\author[Irvine]{Steven R. White}
\address[Marburg]{Fachbereich Physik, Philipps--Universit\"at Marburg,
D--35032 Marburg, Germany}
\address[Irvine]{Department of Physics and Astronomy,  University of
  California, Irvine CA 92697-4575, USA}
\corauth[ich]{Fax:++49-6421-28-24511}
\date{{\bf August 22, 2005}}

\begin{abstract}
Quantum information theory gives rise to a straightforward
definition of the interaction of electrons $I_{p,q}$ in two
orbitals $p$, $q$ for a given many-body wave function. A
convenient way to calculate the von Neumann entropies needed is
presented in this work, and the orbital interaction $I_{p,q}$ is
successfully tested for different types of chemical bonds. As an
example of an application of $I_{p,q}$ beyond the interpretation
of wave functions, $I_{p,q}$ is then used to investigate the
ordering problem in the density-matrix renormalization group.

\end{abstract}
\end{frontmatter}

\section{Introduction}
\label{sec:introduction}

Orbital interaction is used mainly as a \textit{qualitative}
concept in chemistry. Examples are frontier orbital theory or
the isolobal principle. Two orbitals are said to interact when
they have similar energy and/or matching spatial distribution
and/or matching occupation. This concept is also
important in theoretical chemistry: for example, in
complete-active-space calculations, one chooses
(interacting) orbitals to form the active space.
However, there is no unique
quantitative measure of orbital interaction.

One way to obtain  a quantitative measure of orbital interaction
is to utilise the concept of entanglement. Namely, if one divides
the Hilbert space for a given correlated wave function into two
parts, then one can determine the entanglement of the two parts
for the given wave function. For example, a division of a Hilbert
space into one orbital and the space spanned by all the other
orbitals defines the entanglement of this specific orbital with
the rest of the Hilbert space. The more an orbital is entangled,
the more it exchanges information with the other orbitals within
the wave function. It is therefore natural to identify
entanglement with orbital interaction. The advantage of this
interpretation is that quantum information
theory~\cite{Peres,Preskill} defines a \textit{quantitative}
measure of the entanglement and hence of orbital interaction: the
von Neumann entropy $S$. The disadvantage of using such a
definition in calculations is that $S$ is a function of
density-matrix eigenvalues. Consequently, a correlated wave
function must be obtained before a von Neumann entropy can be
calculated.

In order to proceed, one must use a method that produces an
approximate wave function from which the von Neumann entropies can
be calculated in a simple fashion. In addition, the approximate
results should be close to those for the exact wave function. One
suitable technique is to solve the full-configuration-interaction
(FCI) problem using the density-matrix renormalization group
(DMRG)~\cite{White1}. This method produces wave functions and
energies with a well-defined error. In addition to control of the
accuracy of the wave function, the DMRG allows easy access to
operators and expectation values in second-quantised form. We use
this approach in order to calculate the von Neumann entropies in a
straightforward way.

In the following, we argue that a reliable measure of the orbital
interaction between two orbitals can be formed by subtracting the
von Neumann entropy of the two orbitals taken together from the
sum of the individual von Neumann entropies of each orbital. We
then explore the utility of this measure and how their properties
depend on the accuracy of the wave function, we investigate four
example molecules which contain different types of chemical bonds:
LiF, CO, $\rm N_2$, and $\rm F_2$. It turns out that the results
do not depend significantly on the accuracy of the wave function.
It is therefore sufficient to determine the orbital interaction
using a comparatively inexpensive calculation.

We then apply this definition of orbital interaction to a problem
in theoretical chemistry: namely, the ordering problem in the DMRG
algorithm. One ingredient needed to define the DMRG algorithm is
to order the orbitals that span the Hilbert space onto a
one-dimensional lattice. This ordering has a significant influence
on the convergence of the energy. In order to find an ordering
that ensures a good convergence, the literature offers several
approaches~\cite{oers2,head-gordon}, which all have one aspect in
common: they define a measure of orbital interaction and order the
orbitals in a way so that strongly interacting orbitals are near
each other. Therefore, the measure of orbital interaction defined
in this work opens up a new way of approaching this problem.

These topics are treated in the following order: First, an
introduction to the calculation of von Neumann entropies and a
definition of the orbital interaction is given in
Section~\ref{sec:QIT}, which also contains a recipe for
calculating these quantities using expectation values of operators
in second quantisation. Section~\ref{sec:DMRG} presents the
aspects of the DMRG relevant for this work. Subsequently, the
orbital interaction is calculated for a set of test molecules in
Section~\ref{sec:accuracy} in order to test the reliability of the
proposed method. Section~\ref{sec:ordering} then presents results
for the ordering problem in the DMRG. Finally, the main findings
are summarised in Section~\ref{sec:summary}.

\section{Entanglement and orbital interaction}
\label{sec:QIT}

\subsection{Definition of the orbital interaction $I_{p,q}$}

It is useful to divide the Hilbert space of a quantum-mechanical
problem (the ``universe'') into two parts, which we will call  the
``system'' and the ``environment''. If the basis of the system is
described by a complete set of states
$\left\{\left|i\right\rangle\right\}$ and that of the environment
is described by the set $\left\{\left|j\right\rangle\right\}$, the
general wave function $\left|\Psi\right\rangle$ can be written
as~\cite{Feynman}
\begin{equation}
\left|\Psi\right\rangle = \sum_{i,j}C_{i,j} \left|i\right\rangle
\left|j\right\rangle \; .
\label{eq:wavefunction}
\end{equation}
With the help of the density operator for a pure state describing
the total problem
$\hat{\rho}=\left|\Psi\left\rangle\right\langle\Psi\right|$, one
can express the reduced density matrix for the system as
\begin{equation}
\rho^{\rm sys}_{i,i'} = \sum_{j} \rho_{ij,i'j} =
\sum_{j}\left\langle j\right|\left\langle
i\right|\Psi\left\rangle\right\langle\Psi\left|i'\right\rangle
\left|j\right\rangle = \sum_{j,i,i'} C_{i,j}^*C_{i',j} \; .
\label{eq:denmat}
\end{equation}
This is the typical way to set up reduced density matrices. It
requires the partition of the wave function as in
Eq.~(\ref{eq:wavefunction}) as well as the knowledge of the
coefficients $C_{i,j}$.

It is also possible, however, to formulate the problem in a
different way. Suppose one has another, arbitrary basis for the
system, $\left\{\left|n\right\rangle\right\}$, then
\begin{equation}
\rho^{\rm sys}_{n,n'} = \sum_{j} \rho_{nj,n'j} =
\sum_{j}\left\langle j\right|\left\langle
n\right|\Psi\left\rangle\right\langle\Psi\left|n'\right\rangle
\left|j\right\rangle\; .
\end{equation}
With the help of Eq.~(\ref{eq:wavefunction}) one can write
\begin{equation}
\rho^{\rm sys}_{n,n'} = \sum_{j,i,i'} C_{i,j}^*C_{i',j}^{}
\left\langle i'\right|n'\left\rangle\right\langle
n\left|i\right\rangle =
\left\langle\Psi\right|\hat{P}_{n',n}\left|\Psi\right\rangle \; ,
\label{eq:matrixelement}
\end{equation}
where
\begin{equation}
\hat{P}_{n',n}= \sum_j
\left|j\right\rangle\left|n'\right\rangle\left\langle
n\right|\left\langle j\right| \; .
\end{equation}
Instead of a sum over the coefficients of the wave function, one
can calculate the elements of the reduced density matrix with the
help of expectation values of the operator $\hat{P}_{n',n}$. The
effect of $\hat{P}_{n',n}$ is to change the state of the system
from $\left| n\right\rangle$ to $\left| n'\right\rangle$.
It can be viewed as a rotation in state space.

After diagonalisation of $\rho^{\rm sys}$ one obtains its
eigenvalues $\omega_\alpha$, which define the von Neumann entropy
of the system~\cite{Peres,Preskill},
\begin{equation}
S^{\rm sys} = -\sum_\alpha \omega_\alpha \ln\omega_\alpha\; .
\label{eq:entropy}
\end{equation}
This quantity describes how much the system is entangled with the
environment for a given
wave function $\left|\Psi\right\rangle$. When the basis
$\left\{\left|n\right\rangle\right\}$ describes only one orbital
$p$, then the system contains only this orbital $p$ and the
corresponding entropy $S^p$ is the
one-orbital entropy.
If the system contains two
orbitals $p$, $q$, then $S^{pq}$ is a two-orbital entropy. Since
the two-orbital system is built up from two subsystems, namely, the
orbitals $p$ and $q$, one can apply the subadditivity property of
$S$:
\begin{equation}
S^{pq} \leq S^p + S^q\; ,
\end{equation}
where the equality holds when $p$ and $q$ are not entangled. The
interpretation is straightforward: $S^p$ describes the
entanglement of $p$ and $S^q$ the entanglement of $q$ with the
rest of the environment, while $S^{pq}$ describes the entanglement of
$p$ \textit{and} $q$ with the rest of the environment. Any entanglement
between $p$ and $q$ reduces $S^{pq}$ with respect to the sum of
$S^p$ and $S^q$. Therefore, one can define the entanglement
between two individual orbitals by
\begin{equation}
I_{p,q} =\frac{1}{2} \left(S^p + S^q - S^{pq}\right)(1-\delta_{pq}) \geq 0
\;, \label{eq:interaction}
\end{equation}
where the Kronecker $\delta$ ensures that $I_{p,p}=0$, and the factor
$1/2$ prevents interactions from being counted twice. The
quantity $I_{p,q}$ is interpreted in the remaining part of this
work as a measure of the orbital interaction.

Although $I_{p,q}$ describes the entanglement between two orbitals, it
is not possible to use $I_{p,q}$ in order to build up entropies of
larger systems. For example, when the system contains the orbitals $p$
and the environment the orbitals $q$, then  
\begin{equation}
S^{\rm sys} > \sum_{p\in{\rm sys}, q\in{\rm env}} I_{p,q} \; 
\label{eq:additive-entropies}
\end{equation}
for all cases that we have investigated. The difference can amount to
$60$\% of $S^{\rm sys}$. 

\subsection{A recipe for the calculation $S$}
\label{sec:QIT-recipe}

In order to calculate an orbital interaction using
Eq.~(\ref{eq:interaction}), one has to determine and diagonalise
the one- and two-orbital reduced density matrices $\rho^p$ and
$\rho^{pq}$. Their matrix elements are calculated in this work with the
help of the right-hand side of Eq.~(\ref{eq:matrixelement}), in which the
matrix elements are written as
expectation values of $\hat{P}_{n',n}$ with respect to the wave function
$\left|\Psi\right\rangle$, where the operator 
changes the state of the system from $\left| n\right\rangle$ to
$\left| n'\right\rangle$.

One suitable way to represent the system basis
$\left\{\left|n\right\rangle\right\}$ is the occupation-number
representation. In particular, if the system consists of only one
orbital $p$,
\begin{equation}
\left\{\left|n\right\rangle\right\} = \left\{
\hat{c}_{p,\uparrow}^\dagger\left|0\right\rangle ,
\hat{c}_{p,\downarrow}^\dagger\left|0\right\rangle ,
\left|0\right\rangle ,
\hat{c}_{p,\uparrow}^\dagger\hat{c}_{p,\downarrow}^\dagger\left|0\right\rangle
\right\}
=\left\{\left|\uparrow\right\rangle ,\left|\downarrow\right\rangle
, \left|0\right\rangle ,
\left|\uparrow\downarrow\right\rangle\right\}
\label{eq:one-orbital-basis}
\end{equation}
where $\hat{c}_{p,\sigma}^\dagger , \hat{c}_{p,\sigma}^{}$ are
creation and annihilation operators for electrons with spin
$\sigma$ in the orbital $p$, and $\left|0\right\rangle$ is the
vacuum state. Eq.~(\ref{eq:one-orbital-basis}) leads to a
one-orbital density matrix of dimension four. If the system
contains two orbitals, then $\left\{\left|n\right\rangle\right\}$
consists of sixteen basis states, where the orbitals $p$ and $q$
are occupied by zero to four electrons
\begin{equation}
\left\{\left|n\right\rangle\right\} = \left\{
\left|\uparrow,0\right\rangle
,\left|\uparrow,\uparrow\right\rangle ,
\left|\uparrow,\downarrow\right\rangle ,
\left|\uparrow,\uparrow\downarrow\right\rangle, \left|\downarrow,
\uparrow\downarrow\right\rangle,\cdots,\left|\uparrow\downarrow,\uparrow\downarrow\right\rangle
\right\} \; , \label{eq:two-orbital-basis}
\end{equation}
which leads to a $16\times16$ two-orbital density matrix.

This representation determines not only the dimension of the resulting
density matrices, but has two additional consequences. The first
is that the matrix elements of the reduced density matrix
$\rho^{\rm sys}_{n,n'}$ are zero if $\left|n\right\rangle$ and
$\left|n'\right\rangle$ differ in the number of electrons or in
the z-component of the spin, for example
\begin{equation}
\rho^{p}_{\left|\uparrow\right\rangle,\left|\uparrow\downarrow\right\rangle}
=
\sum_j\left\langle\Psi\right|j\left\rangle\right|\uparrow\downarrow\left\rangle\right\langle
\uparrow\left|\left\langle j\right|\Psi\right\rangle = 0 \; .
\end{equation}
Since all of the basis states in Eq.~(\ref{eq:one-orbital-basis})
differ in either the number of electrons or the z-component of the
spin, the one-orbital density matrix is diagonal in this
representation, as shown in Table \ref{tab:one-orbital-denmat}. The
second consequence of Eqs.~(\ref{eq:one-orbital-basis})
and~(\ref{eq:two-orbital-basis}) is that the matrix elements
$\left\langle\Psi\right|\hat{P}_{n',n}\left|\Psi\right\rangle$ of
Eq.~(\ref{eq:matrixelement}) can be expressed using creation and
annihilation operators. For example,
\begin{eqnarray}
\rho^{pq}_{\left|\uparrow,\downarrow\right\rangle,\left|0,\uparrow\downarrow\right\rangle}
&=&
\sum_j\left\langle\Psi\right|j\left\rangle\right|\uparrow,\downarrow\left\rangle\right\langle
0,\uparrow\downarrow\left|\left\langle j\right|\Psi\right\rangle 
\nonumber \\
&=&\left\langle\Psi\left|\hat{c}_{p,\uparrow}^\dagger
\left(1-\hat{n}_{p,\uparrow}^{}\right)\cdot\hat{n}^{}_{q,\downarrow}\hat{c}^{}_{q,\uparrow}
\right|\Psi_{}^{}\right\rangle \; , \label{eq:matrixelement2}
\end{eqnarray}
where
$\hat{n}_{p,\sigma}=\hat{c}_{p,\sigma}^\dagger\hat{c}_{p,\sigma}^{}$
is the (occupation-)number operator for electrons with spin $\sigma$ in
the orbital $p$.
We exploit this property in order to determine the one- and
two-orbital density matrices.

\begin{table}[ht]
\begin{center}
\caption{The one-orbital density matrix $\rho^p_{n,n'}$ for an
orbital $p$. The states $\{\left|n\right\rangle\}$,
$\{\left|n'\right\rangle\}$ are defined in
Eq.~(\ref{eq:one-orbital-basis}). The (occupation-)number operator
for electrons with spin $\sigma$ in the orbital $p$ is
 $\hat{n}_{p,\sigma}=\hat{c}_{p,\sigma}^\dagger\hat{c}_{p,\sigma}$.
\label{tab:one-orbital-denmat}}
\begin{tabular}{l|cccc}\hline
$\rho^p_{n,n'}$ & $\left\langle\uparrow\right|$ &
$\left\langle\downarrow\right|$ & $\left\langle 0\right|$ &
$\left\langle\uparrow\downarrow\right|$ \\ \hline
$\left\langle\uparrow\right|$ & $\left\langle
\hat{n}_{p,\uparrow}\left(1-\hat{n}_{p,\downarrow}\right)\right\rangle$
& 0 & 0 & 0 \\
$\left\langle\downarrow\right|$ & 0 & $\left\langle
\hat{n}_{p,\downarrow}\left(1-\hat{n}_{p,\uparrow}\right)\right\rangle$
& 0 & 0 \\
$\left\langle 0\right|$ & 0 & 0 & $\left\langle\left(1-
\hat{n}_{p,\downarrow}\right)\left(1-\hat{n}_{p,\uparrow}\right)\right\rangle$
& 0 \\
$\left\langle\uparrow\downarrow\right|$ & 0 & 0 & 0 &
$\left\langle
\hat{n}_{p,\downarrow}\hat{n}_{p,\uparrow}\right\rangle$
\\ \hline
\end{tabular}
\end{center}
\end{table}

Consequently, the eigenvalues of $\rho^p_{n,n'}$ in
Table~\ref{tab:one-orbital-denmat} are written in terms of the
occupation-number operator. Thus, the resulting one-orbital entropy $S^p$ depends only on the
occupancy of the orbital. In particular, $S^p = 0$ when the
orbital is fully occupied or empty, and it reaches its maximal
value $S^p_{\rm max}=\ln 4$ when
$\left\langle\hat{n}_{p,\downarrow}\right\rangle =
\left\langle\hat{n}_{p,\uparrow}\right\rangle = 0.5$ and
$\left\langle\hat{n}_{p,\downarrow}\hat{n}_{p,\uparrow}\right\rangle =
0.25$. 

\begin{figure}[ht]
  \begin{center}
    \includegraphics[width=8cm]{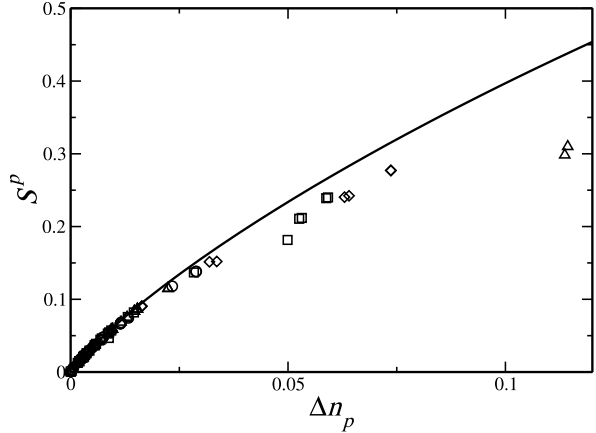}
  \end{center}
\caption{One-orbital entropy $S^p$ as defined in Eq.~(\ref{eq:entropy})
  and Table~~\ref{tab:one-orbital-denmat} versus the deviation of the
  occupancy of orbital $p$, $ n_p =
\left\langle\hat{n}_{p,\downarrow}\right\rangle +
\left\langle\hat{n}_{p,\uparrow}\right\rangle$, from the
  Hartree-Fock case ($\Delta n_p = 2-n_p$ if $n_p\geq 1$ and $\Delta n_p =
  n_p$ if $n_p<1$): circles - LiF, squares - CO, diamonds - $\rm N_2$,
  triangles - $\rm F_2$. Solid line - Eq.~(\ref{eq:Sp-approx}). The values of $S^p$ are taken from a DMRG
  calculation at $m=200$ and Hartree-Fock ordering.} \label{fig:Sp-vs-occ}
\end{figure}

In this work, the occupancies of the orbitals are the same for each spin $\left\langle\hat{n}_{p,\downarrow}\right\rangle =
\left\langle\hat{n}_{p,\uparrow}\right\rangle$, and, using the
approximation
$\left\langle\hat{n}_{p,\downarrow}\hat{n}_{p,\uparrow}\right\rangle
\approx \left\langle\hat{n}_{p,\downarrow}\right\rangle 
\left\langle\hat{n}_{p,\uparrow}\right\rangle$, one gets
\begin{equation}
S^p(n_p) \approx 2\left[
\left(1-\frac{n_p}{2}\right)\ln\left(1-\frac{n_p}{2}\right) + 
\frac{n_p}{2}\ln\frac{n_p}{2}
\right] \; , \label{eq:Sp-approx}
\end{equation}
where $n_p=\left\langle\hat{n}_{p,\downarrow}\right\rangle +
\left\langle\hat{n}_{p,\uparrow}\right\rangle$ is the occupancy of
orbital $p$. Thus, $S^p(n_p)$ is a symmetric function of the occupancy in this case:
$S^p(n_p)=S^p(2-n_p)$. We plot therefore
the values of $S^p$ found in this work versus the deviation $\Delta
n_p$ of the occupancy from the Hartree-Fock case in
Fig.~\ref{fig:Sp-vs-occ}. One to six orbitals deviate from the
approximation of Eq.~(\ref{eq:Sp-approx}), depending on the molecule. The deviations occur for $\Delta n_p\geq
0.025$. The largest one-orbital entropy found in this work $S^p\leq
0.31$ is much
smaller than the maximal value $S^p_{\rm max}\approx 1.38$ because
the occupancies do not deviate strongly from the Hartree-Fock case
and because $\left\langle\hat{n}_{p,\downarrow}\hat{n}_{p,\uparrow}\right\rangle
\neq \left\langle\hat{n}_{p,\downarrow}\right\rangle 
\left\langle\hat{n}_{p,\uparrow}\right\rangle$.

Finally, in this representation it is possible to see that the
two-orbital density matrices also contain the information
contained in the reduced one-particle density matrix
\begin{equation}
\rho^{\rm one}_{p,q} =  \sum_{\sigma} \left\langle \Psi
\left|\hat{c}^{\dagger}_{p,\sigma}\hat{c}^{}_{q,\sigma}
\right|\Psi\right\rangle  \label{eq:one-particle-denmat}
\end{equation}
and the reduced
two-particle density matrix
\begin{equation}
\rho^{\rm two}_{p,q} = \sum_{\sigma,\sigma '} \left\langle \Psi
\left|\hat{n}_{p,\sigma}\hat{n}_{q,\sigma'}
\right|\Psi\right\rangle \; .
\end{equation}
The elements of $\rho^{\rm one}$ are contained in the off-diagonal
elements of the two-orbital density matrices $\rho^{pq}$ while the
elements of $\rho^{\rm two}$ are contained in those diagonal
elements of $\rho^{pq}$ where $p$ and $q$ are always occupied.

This leads to the following recipe to calculate the orbital
entanglement of Eq.~(\ref{eq:interaction}):
\begin{enumerate}
\item Calculate an approximate wave function $|\Psi\rangle$.
\item Calculate the one-orbital and two-orbital density matrices
$\rho^p$ for every orbital $p$ and $q$, and $\rho^{pq}$ for every
combination $p$, $q$.
\item Diagonalise the density matrices and calculate the entropies
  $S^p$ and $S^{pq}$ using the eigenvalues $\omega_\alpha$.
\item Calculate $I_{p,q}$.
\end{enumerate}
In order to follow this recipe and to make use of the formulation
of Eqs.~(\ref{eq:matrixelement}) and (\ref{eq:matrixelement2}), it
is necessary to apply a method that can easily calculate
expectation values of operators in second quantisation with
respect to the total wave function. This is the case for the DMRG,
which is described in the following section.

\section{The DMRG and the calculation of von Neumann entropies}
\label{sec:DMRG}

This section presents only those aspects of the DMRG that are
relevant for the following discussion. For further details we
refer the reader to Refs.~\cite{White1,the-book,schollwoeck}.

\subsection{The setup}

The DMRG is used here to determine the ground state and its energy
for the time-independent, non-relativistic, electronic Hamiltonian
(Full-CI problem) with a well-defined error. The second-quantised
Hamiltonian reads
\begin{equation}
\hat{H}=\sum_{p,q,\sigma} T^{\sigma}_{p,q}\;
\hat{c}^{\dagger}_{p,\sigma}\hat{c}^{}_{q,\sigma}+
\sum_{p,q,r,s,\sigma,\sigma'}V^{\sigma,\sigma'}_{p,q,r,s}\;
\hat{c}^{\dagger}_{p,\sigma}\hat{c}^{\dagger}_{q,\sigma'}
\hat{c}^{}_{r,\sigma'}\hat{c}^{}_{s,\sigma} \; .
\label{eq:Hamiltonian}
\end{equation}
Typically, $T^{\sigma}_{p,q}$, $V^{\sigma,\sigma'}_{p,q,r,s}$ are
the one- and two-electron integrals in the basis of the canonical
orbitals $p,q,r,s$, i.e., the eigenfunctions of the Fock-operator.

The determination of \textit{all} of the eigenstates of
Eq.~(\ref{eq:Hamiltonian}) requires a diagonalisation in a Hilbert
space which grows exponentially with the number of orbitals $N$:
${\rm dim}(\mathcal{H})= 4^{N}$. The key idea of the DMRG is to
form a reduced basis in an optimal, controlled way for the
\textit{few} eigenstates one is interested in, thus reducing the
numerical effort. The size of the reduced basis $\mathcal{B}$ in
this work is ${\rm dim}(\mathcal{B})= 16\cdot m^2$.

\begin{figure}[ht]
  \begin{center}
    \includegraphics{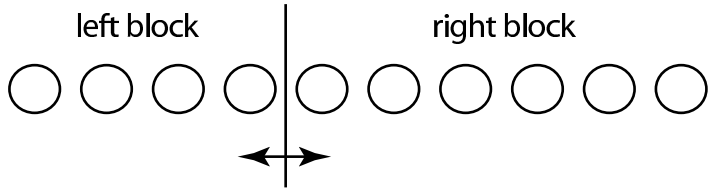}
  \end{center}
\caption{Setup of the DMRG: every orbital occupies one lattice
site. In every step of the DMRG the separator is moved one
site to the right or to the left.} \label{fig:lattice}
\end{figure}

For a given $m$, one determines the optimal basis iteratively with
the help of the reduced density matrix (Eq.~(\ref{eq:denmat})).
This requires the formation of a wave function as in
Eq.~(\ref{eq:wavefunction}) in every step. Consequently, one also must divide
the total basis into a ``system'' and an ``environment'' part
which is realised in the following way: First, one orders the
orbitals onto a one-dimensional lattice, see
Fig.~\ref{fig:lattice}. Second, a boundary line defines a left
and a right block. The product states that can be formed with the
orbitals of each block are represented in this work by $4m$
states: the states of one block are the ``system'' states
$\left\{\left|i\right\rangle\right\}$ and the states of the other
block are the ``environment'' states
$\left\{\left|j\right\rangle\right\}$ in
Eqs.~(\ref{eq:wavefunction}) and~(\ref{eq:denmat}).

In every step of the algorithm the boundary is moved by one site. When
the separator is moved from left to right, the left block
plays the role of the ``system'' and the right block the role of the
``environment''. When the separator is moved 
from right to left the roles are exchanged. A zipper-like motion of the
boundary back and forth through the whole lattice is called a
sweep.

Finally, it is important to point out that the ordering of the
orbitals onto the lattice is in principle arbitrary. One can
therefore choose to order the orbitals on the lattice with
increasing energy, an ordering which we will call the Hartree-Fock
(HF) ordering for the rest of this work. The influence on the
energy convergence of different orderings, as mentioned in
Section~\ref{sec:introduction}, is discussed in
Section~\ref{sec:ordering}.

In the subsequent sections, the ground states of LiF, CO, $\rm
N_2$, and $\rm F_2$ are investigated. The same cc-pVDZ basis
set~\cite{basis} has been used for all molecules, which gives rise
to $N=28$ canonical orbitals. We have calculated the one- and
two-electron integrals using Dalton, a standard quantum-chemistry
program package~\cite{DALTON}. The molecules have been calculated
at their experimental distances: $r_{\rm LiF} = 156.3864$~pm,
$r_{\rm CO} = 112.8323$~pm, $r_{\rm N2} = 109.768$~pm, $r_{\rm F2}
= 141.193$~pm~\cite{LiF,CO,N2,F2}.

\subsection{The error in energy and the measurement of $I_{p,q}$}

The states $\left|i\right\rangle$ of the system block are
improved in every step of the algorithm. In particular, one
diagonalises the Hamiltonian (Eq.~(\ref{eq:Hamiltonian})) in the
reduced basis $\mathcal{B}$, calculates the reduced density matrix
$\rho^{\rm sys}$ (Eq.~(\ref{eq:denmat})) and projects the $4m$
basis states representing the system block onto those $m$
eigenstates of $\rho^{\rm sys}_{i,i'}$ that have the largest
eigenvalues $\omega_\alpha$ and are in that sense an optimal
representation of the system block. The eigenvlues $\omega_\alpha$ also define
the projection error of one step of the DMRG
\begin{equation}
P_m^{\rm step} = 1 - \sum_{\alpha=1}^{m} \omega_\alpha
\end{equation}
since $\sum_\alpha \omega_\alpha =1$. We take the projection error
for a given $m$, $P_m$, to be $P_m^{\rm step}$ at the step when
the separator is in the middle of the lattice in the last sweep,
assuming that convergence in the number of sweeps has been
achieved.

If $m$ is large enough so that
${\dim}\mathcal{B}={\dim}\mathcal{H}$, then the Full-CI
problem is solved and $P_m = 0$. Since the DMRG is variational,
every energy for a given $m$ is larger than the exact energy $E_m
\geq E_{\rm exact}$. Therefore, $E_m$ can be extrapolated to the exact
(Full-CI) ground state
energy $E_{\rm exact}$, which is in our case unknown,
using~\cite{head-gordon}
\begin{equation}
E_m  = E_{\rm exact} + \alpha P_m \; .
\label{eq:extrapolation}
\end{equation}
In order to obtain $E_{\rm exact}$, we calculate a linear
regression to $E_m(P_m)$ at six different $m=(200, 300, 400, 500,
600)$. Typically, six sweeps are necessary to obtain a converged
energy at $m=200$ and four additional sweeps for each of the other values of $m$.
Table \ref{tab:data} contains the extrapolated energies $ E_{\rm
exact}$. The errors given in Table \ref{tab:data} are the standard
deviations for $ E_{\rm exact}$ due to the extrapolation
procedure, and we use error in this sense for the rest of this
manuscript. There are other definitions in the literature for the
error in energy which also depend on $P_m^{\rm step}$~\cite{oers1}
and which consequently should give similar results.

In order to calculate the orbital interaction $I_{p,q}$, one has to
 form the operators that lead to the reduced one- and two-orbital
 density matrices using Eqs.~(\ref{eq:matrixelement}) and~(\ref{eq:two-orbital-basis}). For example,
 instead of evaluating Eq.~(\ref{eq:matrixelement2}) as a whole, one
 determines
\begin{equation}
\rho^{pq}_{\left|\uparrow,\downarrow\right\rangle,\left|0,\uparrow\downarrow\right\rangle}
=
\left\langle\Psi\left|\hat{c}_{p,\uparrow}^\dagger\cdot\hat{n}^{}_{q,\downarrow}\hat{c}^{}_{q,\uparrow}
\right|\Psi_{}^{}\right\rangle - \left\langle\Psi\left|\hat{c}_{p,\uparrow}^\dagger\hat{n}_{p,\uparrow}^{}\cdot\hat{n}^{}_{q,\downarrow}\hat{c}^{}_{q,\uparrow}
\right|\Psi_{}^{}\right\rangle \; . \label{eq:DMRG-operators} 
\end{equation}
In this way, every matrix element of $\rho^{pq}_{n,n'}$ is expressed
as sum of matrix elements of smaller operators. Only 23 of them are
needed to construct  $\rho^{pq}_{n,n'}$. They are stored in
matrix form and must be transformed into the new basis at every step
of the DMRG. We evaluate the 23 operators at the end of a DMRG
calculation with the wave function $\left|\Psi\right\rangle$ from the
last step and calculate $\rho^{pq}_{n,n'}$ from the DMRG output.

Although the one-orbital entropies can be determined throughout
the DMRG algorithm by calculating the local densities
$\left\langle \hat{n}_{i,\sigma} \right\rangle$ via
Table \ref{tab:one-orbital-denmat}, it is necessary to also calculate
them at the end of the procedure in order to obtain comparable
accuracy in each matrix element and in order to calculate them with respect to
the same wave function as the two-orbital entropies.

\section{The structure and accuracy of $I_{p,q}$}
\label{sec:accuracy}

We have carried out calculations for the electronic ground state
of the four test molecules (LiF, CO, $\rm N_2$, $\rm F_2$).
We have obtained wave functions with different energies and errors by
changing some parameters of the algorithm: the
underlying symmetry of the Hartree-Fock calculation, the ordering
of the orbitals, and the parameter $m$. It turns out that the
general structure of $I_{p,q}$ is not affected by these
parameters. In other words, $I_{p,q}$ can already be determined
using a comparatively inexpensive calculation, for example with
$m=200$ at HF ordering in any symmetry.

\begin{figure}[ht]
  \begin{center}
    \includegraphics[width=12cm]{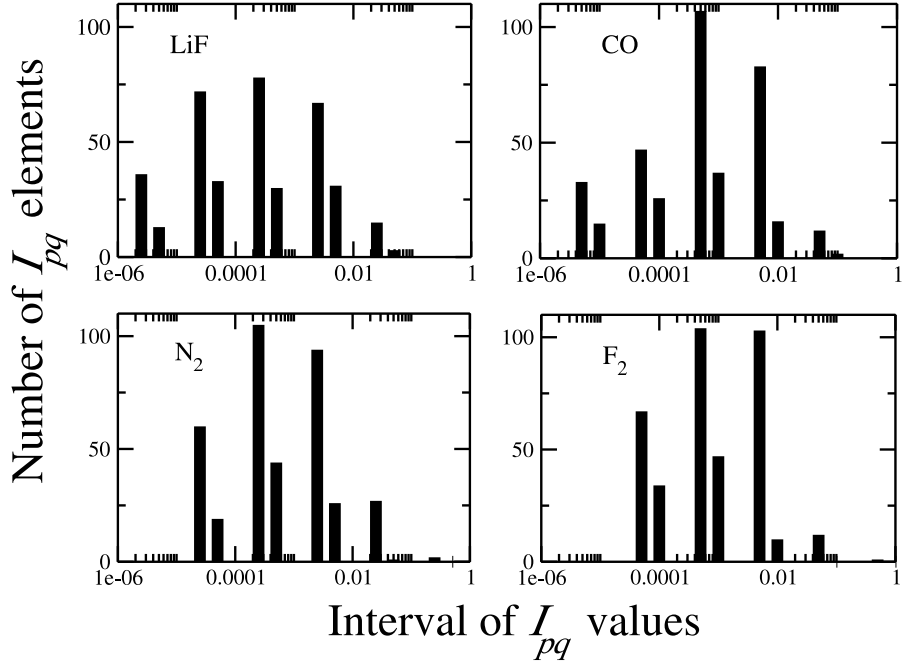}
  \end{center}
\caption{Histograms of $I_{p,q}$ calculated at HF ordering for
$m=200$: The number of $I_{p,q}$ elements are counted for consecutive
 intervals. For example, the number of elements in $0.0001\leq I_{p,q}<0.0005$ is
displayed at $0.0005$, the number in $0.0005\leq I_{p,q}<0.001$ is
displayed at $0.001$ and so on.}\label{fig:histogram}
\end{figure}
\begin{figure}[ht]
  \begin{center}
    \includegraphics[width=12cm]{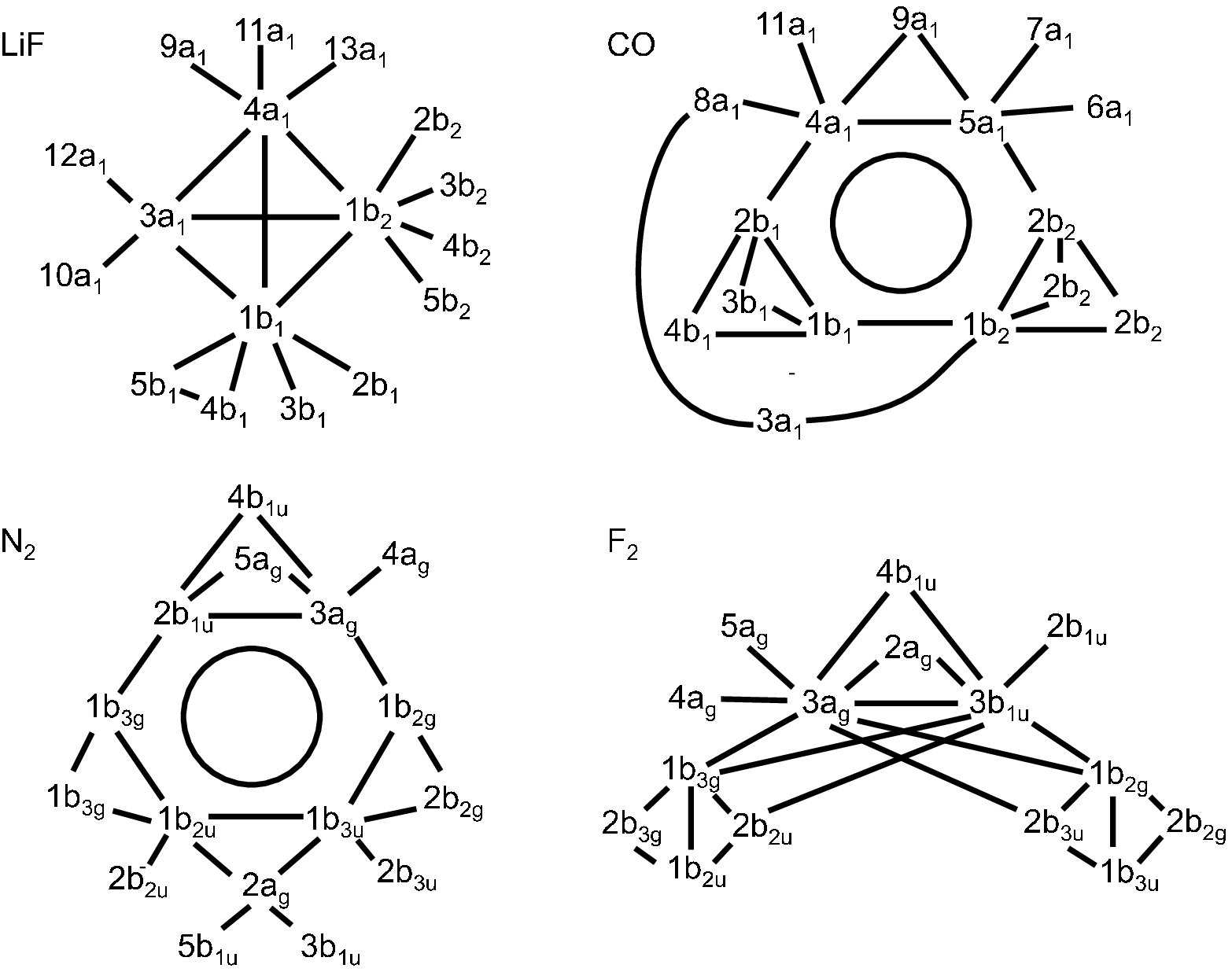}
  \end{center}
\caption{Diagram of $I_{p,q}$ calculated at HF ordering and
$m=200$: Lines connect orbital labels with $I_{p,q}>0.01$. The
circle for CO and $\rm N_2$ denotes that the surrounding orbitals
are all connected with each other.}\label{fig:interaction}
\end{figure}

The values of $I_{p,q}$ are in the range
$0<I_{p,q}<0.18$ in this work. The maximal values for the different molecules are
$I_{p,q}=0.03$ for LiF, $I_{p,q}=0.08$ for CO, $I_{p,q}=0.11$ for $\rm
N_2$, and $I_{p,q}=0.18$ for $\rm F_2$. The distribution of the
$I_{p,q}$ values is shown in Fig.~(\ref{fig:histogram}). One can see
that LiF, the only molecule in the series with an ionic bond, has
a broad distribution of elements, while the other molecules show a
somewhat smaller distribution and two or three intervals with a large
concentration of $I_{p,q}$ elements. The largest interaction $I_{p,q}$
for $\rm F_2$ is one order of magnitude larger than the second largest
interaction element. Also $\rm N_2$ and CO have two to three
interaction elements that are clearly separated from the rest, which
is not the case for LiF. 

In order to visualise the structure of $I_{p,q}$, it is useful to
assign a label to each orbital. In this work, the orbital labels
stem from a Hartree-Fock calculation using the highest point group
available in the Dalton program package: $\rm D_{2h}$ for $\rm
F_2$, $\rm N_2$ and $C_{2v}$ for LiF, and CO. This means that the
lowest-lying orbital in $\rm F_2$ always has the label $1a_g$,
even for calculations in $C_1$. Using these labels, we can specify
the order in which the orbitals are put on the lattice in the DMRG
calculation. The orderings are given in Tables
\ref{tab:LiF-CO-orderings} and~\ref{tab:N2-F2-orderings}, together with
a label for the different cases. The labels designate the
molecule, the ordering criterion, and the point group in which the
HF calculation is carried out. For example, LiF-HF-$C_1$ denotes a
calculation for the LiF molecule, ordering by the orbital energies
(HF ordering), and utilising a HF calculation in $C_1$.

The structure of $I_{p,q}$ can be examined using diagrams. We
connect two orbital labels with a line if the corresponding value
for $I_{p,q}$ is larger than a chosen threshold, here
$I_{p,q}>0.005$. In Fig.~\ref{fig:interaction}, we display such
diagrams for the four molecules studied where the calculations
have been carried out using the HF ordering and $m=200$. From the
diagrams, one can see the following features: First, in all cases
the interaction couples predominantly orbitals of the same
irreducible representation. Second, there are a few orbitals
which are also coupled to orbitals of other irreducible
representations (for example $\rm 3a_1$, $\rm 4a_1$, $\rm 1b_1$,
$\rm 1b_2$ in LiF). Third, those orbitals which couple different
symmetry sectors are often energetically close to the highest
occupied molecular orbital (HOMO). They have an occupancy
$n_p$ that deviates from $n_p=2$ or $n_p=0$ and thus have a large
one-orbital entropy $S^p$ according to Section~\ref{sec:QIT}. 

Therefore, one can say that orbitals with a large one-orbital entropy
$S^p$ also have large interactions $I_{p,q}$. They are often
energetically close to the HOMO and are frontier orbitals in that
sense (see Section~\ref{sec:introduction}). One can also ask, whether the value $I_{p,q}$ is connected to the integrals
$T^{\sigma}_{p,q}$ and $V^{\sigma,\sigma'}_{p,q,r,s}$ of the
Hamiltonian in Eq.~(\ref{eq:Hamiltonian}). To answer this question, we
have a look at the largest $I_{p,q}$ which always correspond to
orbitals $p-q$ which are $\sigma_z-\sigma_z^*$, $\pi_x-\pi_x^*$, or
$\pi_y-\pi_y^*$ combinations. The bonding and
anti-bonding orbitals belong to the same irreducible representation
for LiF and CO.
For $\rm F_2$ and $\rm N_2$, however, these orbitals belong to different
irreducible representations. Consequently, $T^{\sigma}_{p,q}=0$ for
the latter cases and $p$, $q$ are only connected by two-electron
integrals in the Hamiltonian, whereas for LiF and CO one can find one-
and two-electron integrals. It is therefore not possible to deduce a
clear correspondence between the size of $I_{p,q}$ and the size of the
one- and two-electron integrals.

Another way to visualise the overall structure of $I_{p,q}$ is to
plot this quantity as a matrix for a given ordering. For example,
the element $I_{2,5}$ for case LiF-HF-$C_{2v}$ denotes the
interaction between the $2a_1$ and $1b_1$ orbital. This also makes
it possible to mirror the effect of different orderings: an
ordering which groups strongly interacting orbitals together has
large elements next to the diagonal in the plot of $I_{p,q}$.

\begin{figure}[ht]
  \begin{center}
    \includegraphics[width=8cm]{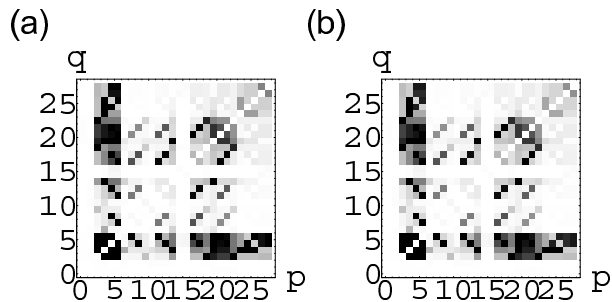}
  \end{center}
\caption{$I_{p,q}$ calculated for ordering LiF-HF-$C_{2v}$ (label
defined in
  Table \ref{tab:LiF-CO-orderings}) with
(a) $m=200$ and (b) $m=600$.} \label{fig:LiF-compare}
\end{figure}

In Fig.~\ref{fig:LiF-compare}, we display plots of $I_{p,q}$ for
the case LiF-HF-$C_{2v}$ with $m=200$ and $m=600$. The increase in
accuracy has no significant effect on $I_{p,q}$, although the
electronic energies differ by about $7\cdot10^{-3}$~a.u. (see
Table \ref{tab:data}). It can also be seen that the HF ordering
results in large weight in the off-diagonal region of $I_{p,q}$
and thus in an ordering for which strongly interacting orbitals
are far apart. This holds also for all other cases of HF ordering.

\begin{figure}[ht]
  \begin{center}
    \includegraphics[width=12cm]{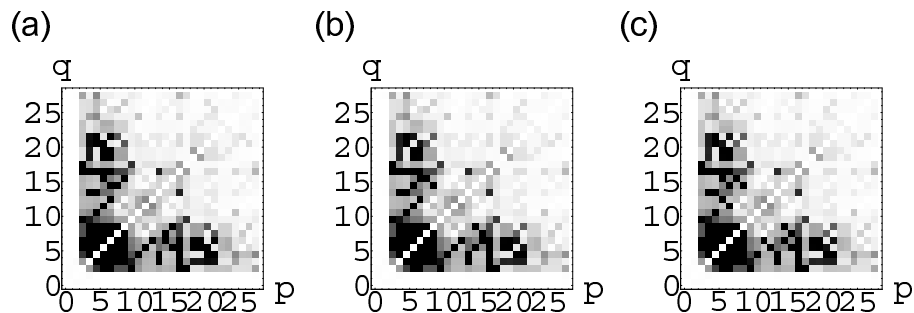}
  \end{center}
\caption{$I_{p,q}$ calculated for (a) $\rm N_2$-HF-$D_{2h}$ and
$m=200$, (b) $\rm N_2$-HF-$C_{2v}$ and $m=600$, and (c) $\rm
N_2$-HF-$C_{1}$ and $m=600$ (labels defined in Table
\ref{tab:N2-F2-orderings}).} \label{fig:N2-compare}
\end{figure}

In Fig.~\ref{fig:N2-compare}, one can see that the influence on
$I_{p,q}$ of different symmetries in the HF calculations is
negligible. The plots of $I_{p,q}$ for $\rm N_2$ and HF ordering
in $D_{2h}$, $C_{2v}$, and $C_1$ are almost identical. This is
also reflected in Fig.~\ref{fig:energies} in which the curves of
the electronic energies with respect to the DMRG steps lie on top
of each other and consequently lead to the same extrapolated
energy (see Table \ref{tab:data}).

\begin{figure}[ht]
  \begin{center}
    \includegraphics[width=12cm]{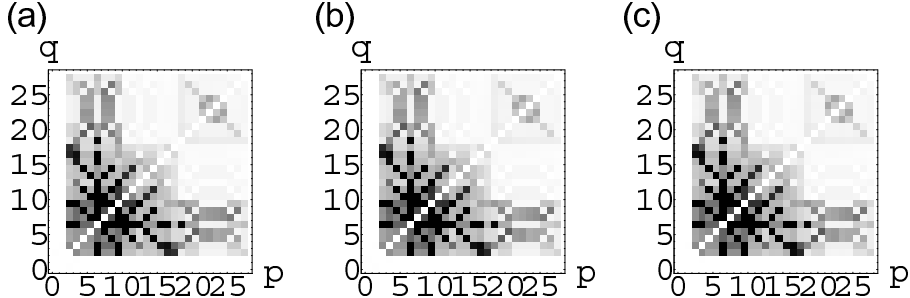}
  \end{center}
\caption{$I_{p,q}$ from calculation using $m=600$ and (a) $\rm
F_2$-HF-$D_{2h}$, (b) $\rm
F_2$-(\ref{eq:sim-an-param-2})-$D_{2h}$, and (c) $\rm
F_2$-\cite{oers3}-$D_{2h}$ (labels defined in Table
\ref{tab:N2-F2-orderings}). The matrices in (b) and (c) are plotted
in the HF ordering rather than the ordering of the actual
calculation.} \label{fig:F2-compare}
\end{figure}

Up to this point, the orbitals have been ordered on the DMRG
lattice according to their energy (HF ordering). Different
orderings result in different wave functions, energies, and
structures of the $I_{p,q}$ matrices. The next section deals with
this issue in more detail. For now, only the effect of the
different orderings on $I_{p,q}$ is discussed.
Fig.~\ref{fig:F2-compare} shows plots of $I_{p,q}$ for $m=600$ and
three cases from Table \ref{tab:N2-F2-orderings}: $\rm
F_2$-HF-$D_{2h}$, $\rm F_2$-(\ref{eq:sim-an-param-2})-$D_{2h}$,
and $\rm F_2$-\cite{oers3}-$D_{2h}$. There is a sizeable energy
difference between the HF ordering and the other two cases of approximately
$6\cdot10^{-3}$~a.u. (see Table \ref{tab:data}). Despite these
differences, no significant difference in the plots of $I_{p,q}$ is
discernible. In order to make this comparison possible, we have
plotted all $I_{p,q}$ matrices in the same ordering, namely the HF
ordering, although the actual calculations have been carried out using
the labelled criteria. In Fig.~\ref{fig:F2-Ipq} one can see plots of
$I_{p,q}$ for $\rm F_2$-(\ref{eq:sim-an-param-2})-$D_{2h}$
and $\rm F_2$-\cite{oers3}-$D_{2h}$ in the ordering which has been
used in the calculation.

To conclude, we find that $I_{p,q}$ is a reliable definition of orbital
interaction which can be calculated for small $m$ in any
ordering. The next section addresses the question of whether one can
use the information in $I_{p,q}$ to obtain an optimal
ordering in the sense that the DMRG has a rapid convergence
towards the exact wave function and energy.

\begin{table}[ht]
\begin{center}
\caption{\textbf{LiF, CO: Orderings of the orbitals as used in the
DMRG:} The labelling has the form
$\langle$molecule$\rangle$-$\langle$ordering
criterion$\rangle$-$\langle$point group$\rangle$ where
$\langle$molecule$\rangle$ is LiF or CO,  $\langle$ordering
criterion$\rangle$ is HF (increasing orbital energy),
(\ref{eq:sim-an-param-2}) (using
Eqs.~(\ref{eq:interaction}),~(\ref{eq:simulated-annealing}),
(\ref{eq:sim-an-param-2})), or \cite{oers3} (Ref.~\cite{oers3}).
The label $\langle$point group$\rangle$ is the corresponding
Sch\"onflies symbol (for example $C_{2v}$). Orbital labels stem
from calculations at the highest possible point group (here
$C_{2v}$). Occupied orbitals are printed in \textbf{bold face}.
\label{tab:LiF-CO-orderings}}
\begin{tabular}{l@{}c@{}c@{}c@{}c@{}c@{}c@{}c@{}c@{}c@{}c@{}c@{}c@{}c@{}c@{}}\hline
Label & \multicolumn{14}{l}{Ordering}\\\hline
LiF-HF-$C_{2v}$ & $\left[\bf 1a_1\right.$ & $\bf 2a_1$ & $\bf
3a_1$ & $\bf 4a_1$ & $\bf 1b_1$ & $\bf 1b_2$ & $5a_1$ & $2b_2$ &
$2b_1$ & $6a_1$ & $7a_1$ & $3b_1$ & $3b_2$ & $8a_1$  \\ & $9a_1$ &
$1a_2$ & $4b_1$ & $4b_2$ & $10a_1$ & $11a_1$ & $5b_1$ & $5b_2$ &
$12a_1$ & $13a_1$ & $6b_2$ & $6b_1$ & $14a_1$ & $\left.
2a_2\right]$ \\
LiF-(\ref{eq:sim-an-param-2})-$C_{2v}$ & 
$\left[ 2b_1\right.$ &  $3b_1$ &  $4b_1$ &  $\bf 1b_1$ &  $5b_1$ &  $14a_1$ &
$2a_2$ & $6b_1$ &  $13a_1$ &  $6b_2$ &  $5a_1$ &  $8a_1$ &  $\bf 4a_1$  &  $11a_1$
\\ &  $12a_1$ &  $\bf 3a_1$ &  $10a_1$ &  $6a_1$ &  $\bf 1a_1$
&  $1a_2$ &  $9a_1$ &  $\bf 2a_1$ &  $7a_1$ &  $2b_2$ &  $3b_2$ &
$4b_2$ &  $\bf 1b_2$ &  $\left. 5b_2\right]$\\
LiF-\cite{oers3}-$C_{2v}$ & $\left[\bf 1a_1\right.$ & $\bf 2a_1$ &
$9a_1$ & $7a_1$ & $5a_1$ & $6a_1$ & $14a_1$ & $13a_1$ & $10a_1$ &
$12a_1$ & $8a_1$ & $\bf 3a_1$ & $11a_1$ & $\bf 4a_1$  \\
 & $2b_1$ & $6b_1$ & $3b_1$ & $4b_1$ & $5b_1$ & $\bf 1b_1$ & $\bf
1b_2$ & $5b_2$ & $4b_2$ & $3b_2$ & $6b_2$ & $2b_2$ & $1a_2$ &
$\left. 1a_2\right]$ \\
CO-HF-$C_{2v}$ & $\left[\bf 1a_1\right.$ & $\bf 2a_1$ & $\bf 3a_1$
& $\bf 4a_1$ & $\bf 1b_1$ & $\bf 1b_2$ & $\bf 5a_1$ & $2b_1$ &
$2b_2$ & $6a_1$ & $3b_2$ & $3b_1$ & $7a_1$ & $8a_1$\\ & $9a_1$ &
$4b_2$ & $4b_1$ & $1a_2$ & $10a_1$ & $5b_1$ & $5b_2$ & $11a_1$ &
$12a_1$ & $13a_1$ & $2a_2$ & $6b_2$ & $6b_1$ &
$\left.14a_1\right]$\\
CO-(\ref{eq:sim-an-param-2})-$C_{2v}$ & $\left[5b_1\right.$ &
$4b_1$ & $\bf 1b_1$ & $2b_1$ & $\bf 5a_1$ &  $7a_1$ &  $3b_1$ &
$1a_2$ & $6b_1$ & $\bf 2a_1$ & $14a_1$ &  $11a_1$ &  $9a_1$ & $\bf
4a_1$
\\ & $8a_1$ & $\bf 3a_1$ &  $12a_1$ &  $6a_1$ &  $\bf 1a_1$ &
$2a_2$ & $13a_1$ & $6b_2$ & $5b_2$ &  $4b_2$ &  $\bf 1b_2$ &
$2b_2$ & $3b_2$ & $\left.10a_1\right]$\\
CO-\cite{oers3}-$C_{2v}$ & $\left[\bf 1a_1\right.$ & $\bf 2a_1$ &
$14a_1$ & $13a_1$ & $10a_1$ & $12a_1$ & $11a_1$ & $6a_1$ & $7a_1$
& $8a_1$ & $9a_1$ & $\bf 3a_1$ & $\bf 4a_1$ & $\bf 5a_1$\\ &
$6b_2$ & $5b_2$ & $3b_2$ & $4b_2$ & $2b_2$ & $\bf 1b_2$ & $\bf
1b_1$ & $2b_1$ & $4b_1$ & $3b_1$ & $5b_1$ & $6b_1$ & $1a_2$ &
$\left.2a_2\right]$\\
\hline
\end{tabular}
\end{center}
\end{table}

\begin{table}[ht]
\begin{center}
\caption{\textbf{$\mathbf N_2$ Orderings of the orbitals as used
in the DMRG:}
 The labelling has the form
$\langle$molecule$\rangle$-$\langle$ordering
criterion$\rangle$-$\langle$point group$\rangle$ where
$\langle$molecule$\rangle$ is $\rm N_2$ or $\rm F_2$,  $\langle$ordering
criterion$\rangle$ is HF (increasing orbital energy),
(\ref{eq:sim-an-param-2}) (using
Eqs.~(\ref{eq:interaction}),~(\ref{eq:simulated-annealing}),
(\ref{eq:sim-an-param-2})), or \cite{oers3} (Ref.~\cite{oers3}).
The label $\langle$point group$\rangle$ is the corresponding
Sch\"onflies symbol (for example $C_{2v}$). Orbital labels stem
from calculations at the highest possible point group (here
$D_{2h}$). Occupied orbitals are printed in \textbf{bold face}.
\label{tab:N2-F2-orderings}}
\begin{tabular}{@{}l@{\hspace{-1.8mm}}c@{}c@{}c@{}c@{}c@{}c@{}c@{}c@{}c@{}c@{}c@{}c@{}c@{}c@{}}\hline
Label& \multicolumn{14}{l}{Ordering}\\\hline
$\rm N_2$-HF-$C_1$ & $\left[\bf 1a_g\right.$ & $\bf 1b_{1u}$ &
$\bf 2a_g$ & $\bf 2b_{1u}$ & $\bf 3a_g$ & $\bf 1b_{3u}$ & $\bf
1b_{2u}$ & $1b_{3g}$ & $1b_{2g}$ & $3b_{1u}$ & $4a_{g}$ &
$2b_{3u}$ & $2b_{2u}$ & $5a_g$\\ $\sim C_{2v}$,$D_{2h}$& $2b_{3g}$
& $2b_{2g}$ & $4b_{1u}$ & $5b_{1u}$ & $6a_g$ & $1b_{1g}$ &
$3b_{3u}$ & $3b_{2u}$ & $6b_{1u}$ & $1a_u$ & $7a_g$ & $3b_{3g}$ &
$3b_{2g}$ & $\left.7b_{1u}\right]$\\
$\rm N_2$-(\ref{eq:sim-an-param-2})-$D_{2h}$& $\left[6a_g\right.$
& $1b_{1g}$ & $2b_{2g}$ & $1b_{2g}$ & $\bf 1b_{3u}$ & $2b_{3u}$ &
$3b_{3u}$ & $1a_u$ & $\bf 1a_g$ & $7b_{1u}$ & $3b_{2u}$ &  $5a_g$
&  $\bf 2b_{1u}$ & $\bf 3a_g$ \\ & $4b_{1u}$ & $\bf 2a_g$ &
$5b_{1u}$ & $3b_{1u}$ & $3b_{2g}$ & $\bf 1b_{1u}$ & $3b_{3g}$ &
$7a_g$ & $4a_g$ & $2b_{2u}$ & $\bf 1b_{2u}$ & $1b_{3g}$ &
$2b_{3g}$ & $\left. 6b_{1u}\right]$
\\
$\rm N_2$-\cite{oers3}-$D_{2h}$ & $\left[\bf 1a_g\right.$ & $7a_g$
& $6a_g$ & $4a_g$ & $5a_g$ &  $\bf 2a_g$ &  $\bf 3a_g$ & $1a_u$ &
$3b_{3g}$ & $2b_{3g}$ &  $1b_{3g}$ &  $1b_{2g}$ & $2b_{2g}$ &
$3b_{2g}$ \\ & $3b_{2u}$ &  $2b_{2u}$ &  $\bf 1b_{2u}$ &  $\bf
1b_{3u}$ & $2b_{3u}$ &  $3b_{3u}$ &  $1b_{1g}$ & $\bf 2b_{1u}$ &
$4b_{1u}$ & $3b_{1u}$ &  $5b_{1u}$ &  $6b_{1u}$ & $7b_{1u}$ &
$\left.\bf 1b_{1u}\right]$ \\
$\rm N_2$-\cite{oers3}-$C_{2v}$ & $\left[\bf 1a_g\right.$ & $\bf
1b_{1u}$ & $7b_{1u}$ &  $6b_{1u}$ &  $7a_g$ & $5b_{1u}$ & $6a_g$ &
$4a_g$ & $3b_{1u}$ &  $5a_g$ &  $4b_{1u}$ & $\bf 2a_g$ & $\bf
3a_g$ & $\bf 2b_{1u}$ \\ & $3b_{2g}$ & $3b_{2u}$ & $2b_{3u}$ &
$2b_{3g}$ & $1b_{2g}$ &  $\bf 1b_{3u}$ & $\bf 1b_{2u}$ & $1b_{3g}$
& $2b_{2g}$ &  $2b_{2u}$ &  $3b_{3u}$ & $3b_{3g}$ & $1b_{1g}$ &
$\left. 1a_u\right]$\\
$\rm N_2$-\cite{oers3}-$C_{1}$ & $\left[\bf 1b_{1u}\right.$ &
$3b_{2g}$ & $6b_{1u}$ &  $7a_g$ &  $1b_{1g}$ & $3b_{2u}$ &
$2b_{2u}$ & $3b_{1u}$ &  $2b_{3g}$ &  $5a_g$ &  $\bf 2a_g$ &  $\bf
2b_{1u}$ & $1b_{3g}$ &  $\bf 1b_{2u}$ \\ &  $\bf 1b_{3u}$ &
$1b_{2g}$ & $\bf 3a_g$ &  $4b_{1u}$ &  $2b_{2g}$ & $2b_{3u}$ &
$4a_g$ & $3b_{3u}$ &  $6a_g$ &  $5b_{1u}$ &  $1a_u$ & $7b_{1u}$ &
$3b_{3g}$ & $\left.\bf 1a_g\right.$ \\
$\rm F_2$-HF-$D_{2h}$ & $ \left[\bf 1a_g\right.$ & $\bf 1b_{1u}$ &
$\bf 2a_g$ & $\bf 2b_{1u}$ & $\bf 1b_{2u}$ & $\bf 1b_{3u}$ & $\bf
3a_g$ & $\bf 1b_{2g}$ & $\bf 1b_{3g}$ & $3b_{1u}$ & $2b_{3u}$ &
$2b_{2u}$ & $4b_{1u}$ & $2b_{2g}$ \\ & $2b_{3g}$ & $4a_g$ & $5a_g$
& $5b_{1u}$ & $6a_g$ & $3b_{2u}$ & $3b_{3u}$ & $7a_g$ & $1b_{1g}$
& $6b_{1u}$ & $1a_u$ & $3b_{2g}$ & $3b_{3g}$ & $\left.
7b_{1u}\right]$\\
$\rm F_2$-(\ref{eq:sim-an-param-2})-$D_{2h}$ & $\left[5a_g\right.$
& $\bf 2b_{1u}$ & $4a_g$ & $\bf 3a_g$ & $3b_{1u}$ & $4b_{1u}$ &
$6a_g$ & $\bf 1b_{1u}$ & $\bf 1a_g$ & $3b_{3g}$ & $3b_{2u}$ &
$2b_{3g}$ & $\bf 1b_{3g}$ & $2b_{2u}$
\\  & $\bf 1b_{2u}$ & $\bf 2a_g$ & $5b_{1u}$ & $7b_{1u}$ & $6b_{1u}$ &
$7a_g$ & $1b_{1g}$ & $1a_u$ & $3b_{2g}$ & $3b_{3u}$ & $\bf
1b_{3u}$ & $2b_{3u}$ & $\bf 1b_{2g}$ &  $\left. 2b_{2g}\right]$\\
$\rm F_2$-\cite{oers3}-$D_{2h}$  & $\left[\bf 1a_g\right.$ &
$7a_g$ & $6a_g$ & $5a_g$ & $4a_g$ &  $\bf 2a_g$ &  $\bf 3a_g$ &
$3b_{1u}$ & $\bf 2b_{1u}$ & $4b_{1u}$ & $5b_{1u}$ &  $6b_{1u}$ &
$7b_{1u}$ & $\bf 1b_{1u}$ \\  & $3b_{2g}$ & $2b_{2g}$ &  $\bf
1b_{2g}$ & $\bf 1b_{3g}$ & $2b_{3g}$ & $3b_{3g}$ & $3b_{3u}$ &
$2b_{3u}$ & $\bf 1b_{3u}$ & $\bf 1b_{2u}$ & $2b_{2u}$ & $3b_{2u}$
&  $1b_{1g}$ & $\left. 1a_u\right]$\\
\hline
\end{tabular}
\end{center}
\end{table}

\begin{table}[ht]
\begin{center}
\renewcommand{\baselinestretch}{1.0}
\caption{\textbf{Energies and errors due to different orderings:}
Label - $\langle$molecule$\rangle$-$\langle$ordering
criterion$\rangle$-$\langle$point group$\rangle$ defined in Tables
\ref{tab:LiF-CO-orderings} and~\ref{tab:N2-F2-orderings}; $m$ - size of
the largest used Hilbert space ${\rm dim}(\mathcal{B})=16\cdot
m^2$; energies - electronic energy in atomic units (without nuclei
interaction), extrapolation according to
Eq.~(\ref{eq:extrapolation}) for $m=200,300,400,500,600$ error is
the standard deviation. \label{tab:data}}
\begin{tabular}{@{}llll}
\hline Label & m & last sweep energy & extrapolated energy (error)
\\\hline
LiF-HF-$C_{2v}$ & 200 & -116.2870771 \\
LiF-HF-$C_{2v}$ & 600  & -116.2936058 & -116.2938841 ($\pm
2\cdot10^{-4}$)\\
LiF-(\ref{eq:sim-an-param-2})-$C_{2v}$ & 600 & -116.2939879
&-116.2940038 ($\pm 2\cdot10^{-5}$)\\
LiF-\cite{oers3}-$C_{2v}$ & 600 & -116.2940057 &-116.2940214 ($\pm
4\cdot10^{-6}$)\\
CO-HF-$C_{2v}$  & 600 &  -135.5668011 &-135.5694985 ($\pm
6\cdot10^{-4}$)\\
CO-(\ref{eq:sim-an-param-2})-$C_{2v}$& 600 &  -135.5697747 &
-135.5711334 ($\pm 7\cdot10^{-4}$)\\
CO-\cite{oers3}-$C_{2v}$ & 600 & -135.5703675 & -135.5709141 ($\pm
1\cdot10^{-4}$)\\
$\rm N_2$-HF-$D_{2h}$  & 600 &  -132.8983153 & -132.9015018 ($\pm
3\cdot10^{-4}$)\\
$\rm N_2$-HF-$C_{2v}$ & 600 & -132.8983171 & -132.9013120 ($\pm
4\cdot10^{-4}$)\\
$\rm N_2$-HF-$C_{1}$ & 600 & -132.8983197 & -132.9014066($\pm
3\cdot10^{-4}$)\\
$\rm N_2$-(\ref{eq:sim-an-param-2})-$D_{2h}$ & 600 & -132.9004719
& -132.9025150 ($\pm 2\cdot10^{-4}$)\\
$\rm N_2$-\cite{oers3}-$D_{2h}$ & 600 & -132.8979579 &
-132.9006192 ($\pm 2\cdot10^{-4}$)\\
$\rm N_2$-\cite{oers3}-$C_{2v}$ & 600 & -132.9020805 &
-132.9027983 ($\pm 4\cdot10^{-5}$)\\
$\rm N_2$-\cite{oers3}-$C_{1}$ & 600 & -132.8966134 & -132.9028483
($\pm 2\cdot10^{-4}$)\\
$\rm F_2$-HF-$D_{2h}$  & 600 & -229.4522256 & -229.4592938 ($\pm
1\cdot10^{-3}$)\\
$\rm F_2$-(\ref{eq:sim-an-param-2})-$D_{2h}$ & 600 & -229.4586797
& -229.4615548 ($\pm 2\cdot10^{-4}$)\\
$\rm F_2$-\cite{oers3}-$D_{2h}$ & 600 & -229.4581772 &
-229.4611454 ($\pm 2\cdot10^{-4}$)\\
\hline
\end{tabular}
\end{center}
\end{table}

\section{The ordering problem}
\label{sec:ordering}

\subsection{The problem and solution strategies}

\begin{figure}[ht]
  \begin{center}
    \includegraphics[width= 13cm]{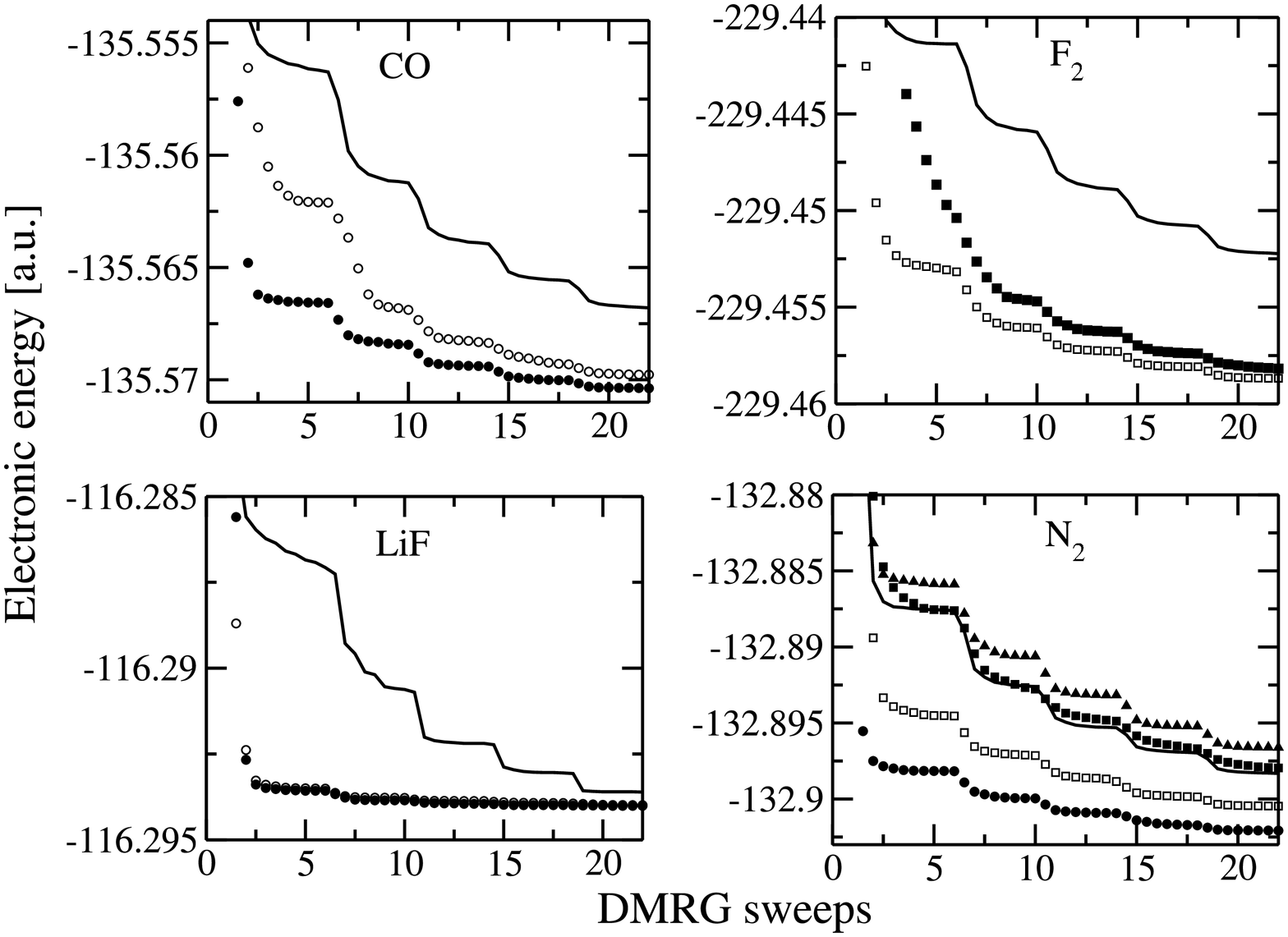}
  \end{center}
\caption{Electronic energy (no nucleus-nucleus interaction) versus
number of DMRG sweeps: $m=200$ (sweeps 1 to 6), $m=300$ (sweeps 7
to 10), $m=400$ (sweeps 11 to 14), $m=500$ (sweeps 15 to 18),
$m=600$ (sweeps 19 to 22): solid line - HF ordering; open symbols
- ordering using
Eqs.~(\ref{eq:interaction}),~(\ref{eq:simulated-annealing}),~(\ref{eq:sim-an-param-2});
filled symbols - ordering using Ref.~\cite{oers3}; circles -
$C_{2v}$; squares - $D_{2h}$; triangles - $C_1$ }
\label{fig:energies}
\end{figure}

Two things must be chosen before a DMRG calculation can be carried
out: the basis and its ordering on the lattice. In principle, any
orthonormal basis of orbitals can be used and the ordering is
arbitrary. However, the ordering does affect the convergence of
the DMRG in practice~\cite{oers2}. This can be seen by analyzing
the behaviour of the energies as $m$ is increased. With increasing
$m$, the variational nature of the DMRG leads to a decrease in
energy for every ordering. For a sufficiently large $m$, the
difference to the exact energy can be made arbitrarily small. In
this sense the ordering is arbitrary. However, the value of $m$
that is necessary for a certain accuracy in energy depends on the
ordering. For example, in Fig.~\ref{fig:energies} one can see that
the case LiF-(\ref{eq:sim-an-param-2})-$C_{2v}$ leads to a much
lower energy for $m=200$ than LiF-HF-$C_{2v}$. With increasing $m$
the energies for LiF-HF-$C_{2v}$ improve and for $m=600$ both
cases show only a difference in energy of $3\cdot 10^{-4}$~a.u., as
seen in Table~\ref{tab:data}. However, the error in the energy in
Table \ref{tab:data} is considerably larger for LiF-HF-$C_{2v}$.
In the worst case, it is also possible to find orderings that
cause a trapping of the DMRG algorithm in a local
minimum~\cite{oers2}. The energy then does not decrease significantly
with increasing $m$ even though $P_m$ remains small. It is important
to note that this trapping is not necessarily due to a certain
ordering but is sometimes caused by an unfavourable choice of parameters
for the DMRG calculation.

In order to apply the DMRG to practical situations, it is
therefore crucial to define a criterion for an optimal ordering
for a given basis, in order to avoid trapping in local minima and
in order to achieve the highest possible accuracy. The approaches
which have been pursued so far have defined an orbital interaction
and have ordered the orbitals so that strongly interacting
orbitals are near each other on the one-dimensional lattice.

One approach is to define the interaction between orbitals $p$ and
$q$ in terms of the one- and two-electron integrals of the
Hamiltonian~\cite{head-gordon,hess}. Improved orderings then
reduce the bandwidth of the $T^\sigma_{p,q}$ matrix, for example.
This approach is not able to avoid trapping in local
minima~\cite{oers2}. In order to find an ordering criterion based
on $T^\sigma_{p,q}$ or $V^{\sigma,\sigma'}_{i,j,i,j}$, a recent
study~\cite{reiher} has investigated a large number of orderings
for the $\rm Cr_2$ molecule using a genetic algorithm, which has
not yet led to a general criterion for different molecules and
basis sets.

\begin{figure}[ht]
  \begin{center}
    \includegraphics[width= 13cm]{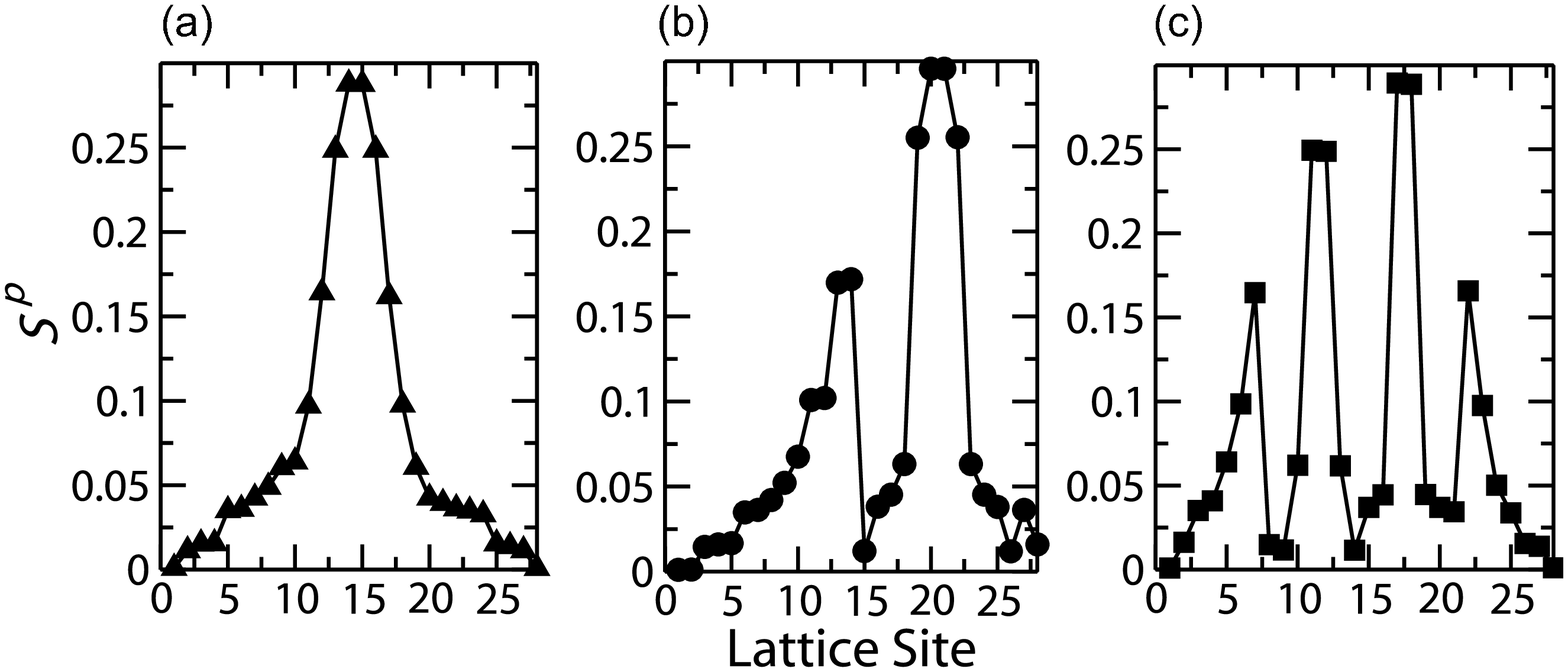}
  \end{center}
\caption{Orbital entropies $S^p$ following Eq.~(\ref{eq:entropy})
and Table \ref{tab:one-orbital-denmat} for (a) $\rm
N_2$-\cite{oers3}-$C_{1}$, (b) $\rm N_2$-\cite{oers3}-$C_{2v}$, and 
(c) $\rm N_2$-\cite{oers3}-$D_{2h}$ (labels defined in Table
\ref{tab:N2-F2-orderings}). } \label{fig:N2-oers-siteent}
\end{figure}

Another approach is to group the orbitals according to their
irreducible representations and then order the orbitals within
these groups in order to maximise the one-orbital entropy $S^p$
along the lattice~\cite{oers2,oers3}.  The net effect is that some
entangled (interacting) orbitals are placed close together but are
also somewhat distributed over the lattice. This is called
``competition between entanglement localisation and interaction
localisation'' in Ref.~\cite{oers3}. The label ``\cite{oers3}'' is
attributed to this criterion, which obviously depends on the
underlying symmetry of the HF calculation. This is illustrated for
$\rm N_2$-\cite{oers3}-$D_{2h}$, $\rm N_2$-\cite{oers3}-$C_{2v}$,
and $\rm N_2$-\cite{oers3}-$C_{1}$ in
Fig.~\ref{fig:N2-oers-siteent}. The distribution of $S^p$ along
the lattice is plotted for the three point groups investigated.
The maxima correspond to boundaries between groups of orbitals
with the same irreducible representation. For the $C_1$ ordering
 in Fig.~\ref{fig:N2-oers-siteent}(a), the sites with large
one-orbital entropy are bunched up in the middle of the lattice.
In Fig.~\ref{fig:energies}, we display the effect of these
orderings on the convergence of the DMRG: only in the case $\rm
N_2$-\cite{oers3}-$C_{2v}$ can one see an improvement over the HF
ordering $\rm N_2$-HF-$D_{2h}$. Therefore, one is not guaranteed
that a new ordering according to the strategy of Ref.~\cite{oers3}
leads to improved energy convergence.

\subsection{New strategy using $I_{p,q}$}

A strategy based on the orbital interaction $I_{p,q}$ combines the
advantages of the two earlier approaches. On the one hand, one is
using the entanglement information of a many-body wave function as
in Ref.~\cite{oers3}. On the other hand, one has a specific
interaction in matrix form as in
Refs.~\cite{head-gordon,hess,reiher}. The quantity $S^p$ is not
sufficiently specific because it only indicates how much orbital
$p$ interacts with all the other orbitals, while $I_{p,q}$ is a
direct measure of the interaction between $p$ and $q$. The
analysis of $I_{p,q}$ in Fig.~\ref{fig:interaction} has also shown
that orbitals from the same irreducible representation are coupled
strongly. It is therefore not necessary to account for this fact
separately as had to be done in Ref.~\cite{oers3}.

Here we search for an improved ordering which localises the
interaction $I_{p,q}$, i.e., reduces the bandwidth of the
$I_{p,q}$ matrix. The optimal ordering is found using simulated
annealing~\cite{numerical-recipes}. We argue that this approach is
an improvement over the Cuthill-McKee algorithm, used for example
in Refs.~\cite{head-gordon,hess}, which distinguishes only between
occupied and unoccupied matrix elements and thus neglects the
information contained in the size of the elements. The annealing
algorithm is comparable to the genetic algorithm~\cite{reiher}
because both probe different, randomly generated configurations.
The cost function in the annealing process plays a similar role to
the fitness function in the genetic algorithm.

Following the idea that good orderings should place strongly
interacting orbitals near each other, we have investigated several cost
functions $F$ that all favour orderings in which large elements of $I_{p,q}$
are on the secondary diagonal and which have a small
bandwidth. It turns out that a cost function
\begin{equation}
F = \frac{I_{p,q}}{r^2}\; ,\label{eq:simulated-annealing}
\end{equation}
where $r=|p-q|$, is a good starting point. In order to increase
the attraction to the secondary diagonal of large elements of
$I_{p,q}$, we set
\begin{equation}
r =
\begin{cases}
0.5    & \text{if } |p-q|=1 \\
|p-q| & \text{otherwise} \; .
\end{cases}
\label{eq:sim-an-param-1}
\end{equation}
We have also made further adjustments to
the cost function $F$. From the discussion of the ordering
criterion in Ref.~\cite{oers3}, we have learned that strongly
interacting orbitals should not be bunched up, (for example, in
the middle of the lattice) but should instead be more evenly
distributed. Therefore, we set $r=0.5$ for elements on the
secondary diagonal in regions of length $N/5$ around the edges and
the middle of the lattice, where $N$ is the number of orbitals,
i.e.,
\begin{equation}
r =
\begin{cases}
0.5    & \text{if $|p-q|=1$ and}\\ & \left(\text{$\{p,q\}\leq N/5$
or } \text{ $\{p,q\}\geq N-N/5$ or}\right.\\ &
\text{$\left.N/2-N/10 \leq\{p,q\}\leq N/2+N/10\right)$}
\\
|p-q| & \text{otherwise} \; .
\end{cases}
\label{eq:sim-an-param-2}
\end{equation}
An ordering created by the use of
Eqs.~(\ref{eq:interaction}),~(\ref{eq:simulated-annealing}) and
(\ref{eq:sim-an-param-2}) will be labelled
``(\ref{eq:sim-an-param-2})''.

In all cases in which orderings have been generated using 
Eq.~(\ref{eq:sim-an-param-2}) we have found an improved energy
convergence with respect to calculations with HF ordering, and the
calculations are not trapped. However, we find that there is no
clear correspondence between the value of the cost function for a
given ordering and the respective energy convergence. In other
words, small changes in the ordering and the cost function can
influence the energy convergence severely. For example, the case
LiF-\cite{oers3}-$D_{2h}$ has a similar energy convergence as the case
LiF-(\ref{eq:sim-an-param-2})-$D_{2h}$  as shown in
Table~\ref{tab:data}. However, $F=0.6397$ in the former case and
$F=1.574$ in the latter, while the HF ordering yields $F=0.281$.

To summarize, while, in our opinion,
Eqs.~(\ref{eq:simulated-annealing}) and~(\ref{eq:sim-an-param-2})
represent an improved cost function, one cannot always be certain
that an ordering determined by these equations really is the
optimal one. While the value of $F$ contains valuable information,
additional information, which can be obtained by the visual
investigation of $I_{p,q}$ as a matrix, is also important. This will be
discussed for specific cases in the following. 

\begin{figure}[ht]
  \begin{center}
    \includegraphics[width=8cm]{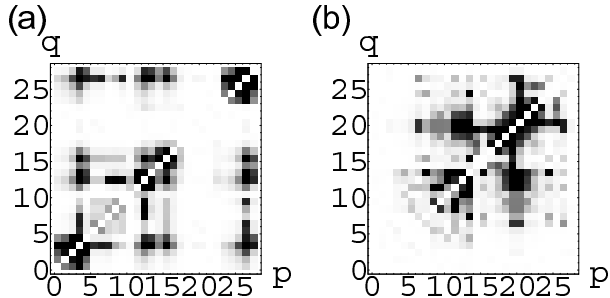}
  \end{center}
\caption{$I_{p,q}$ calculated at $m=600$ for
(a) LiF-(\ref{eq:sim-an-param-2})-$C_{2v}$, and (b)
LiF-\cite{oers3}-$C_{2v}$ (labels defined in Table
\ref{tab:LiF-CO-orderings}).} \label{fig:LiF-Ipq}
\end{figure}

LiF, for example, is a rather unproblematic case: any new ordering
that reduces the bandwidth of $I_{p,q}$ results in improved
convergence in the DMRG. In Fig.~\ref{fig:LiF-Ipq}(a), one can see
that the bandwidth of  $I_{p,q}$ for the ordering of
this work is reduced compared to the HF ordering, which results in
a reduced extrapolated energy and a reduced error (see
Table \ref{tab:data} and Fig.~\ref{fig:energies}). While the result
using the criterion of Ref.~\cite{oers3} looks more compact
(Fig.~\ref{fig:LiF-Ipq}(b)), it yields similar results for the
energy convergence. A comparison of the orbital orderings in
Table \ref{tab:LiF-CO-orderings} shows that both cases group
together orbitals of the same irreducible representation.

\begin{figure}[ht]
  \begin{center}
    \includegraphics[width=8cm]{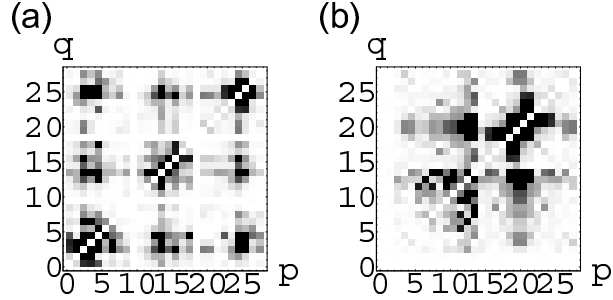}
  \end{center}
\caption{$I_{p,q}$ calculated at $m=600$ for (a)
CO(\ref{eq:sim-an-param-2})-$C_{2v}$, and (b)
CO-\cite{oers3}-$C_{2v}$ (labels defined in Table
\ref{tab:LiF-CO-orderings}).} \label{fig:CO-Ipq}
\end{figure}

For CO and $\rm N_2$, the situation is more complicated. In
Fig.~\ref{fig:CO-Ipq}(a), one can see that the form of $I_{p,q}$ for
our ordering is again more spread out than for the ordering
of Ref.~\cite{oers3} (Fig.~\ref{fig:CO-Ipq}(b)). This
time, however, the latter criterion leads to a more exact
energy, as can be seen in Fig.~\ref{fig:energies}.

\begin{figure}[ht]
  \begin{center}
    \includegraphics[width=12cm]{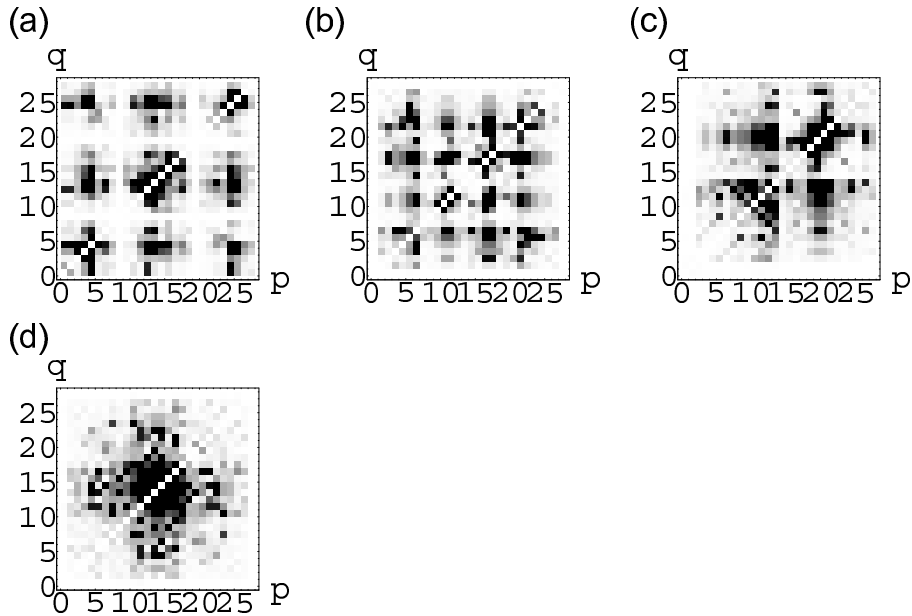}
  \end{center}
\caption{$I_{p,q}$ calculated at $m=600$ for (a) $\rm
N_2$-(\ref{eq:sim-an-param-2})-$D_{2h}$, (b) $\rm
N_2$-\cite{oers3}-$D_{2h}$, (c) $\rm N_2$-\cite{oers3}-$C_{2v}$,
and (d) $\rm N_2$-\cite{oers3}-$C_{1}$ (labels defined in Table
\ref{tab:N2-F2-orderings}).} \label{fig:N2-Ipq}
\end{figure}

Fig.~\ref{fig:N2-Ipq}(a) shows $I_{p,q}$ for $\rm N_2$ plotted for the
ordering criterion of this work.
Figs.~\ref{fig:N2-Ipq}(b)-(d) are determined by the criterion of
Ref.~\cite{oers3} and use different symmetries in the underlying
HF calculation. All four plots have a reduced bandwidth compared
to the HF ordering shown in Fig.~\ref{fig:N2-compare}, but only
the cases in Figs.~\ref{fig:N2-Ipq}(a) and \ref{fig:N2-Ipq}(c)
show better energy convergence. The criterion of Ref.~\cite{oers3}
yields a slightly more compact form of $I_{p,q}$, as can be seen
in Fig.~\ref{fig:N2-Ipq}(c), and leads to a better convergence in
energy than our criterion (see Fig.~\ref{fig:energies}). The cases
displayed in Figs.~\ref{fig:N2-Ipq}(b) and ~\ref{fig:N2-Ipq}(d)
distribute or accumulate the interacting orbitals too much,
leading to poor energy convergence, as explained in the discussion
of Fig.~\ref{fig:N2-oers-siteent}. This underscores that subtle
changes in the structure of the $I_{p,q}$ matrix influence the
energy convergence of the DMRG.

\begin{figure}[ht]
  \begin{center}
    \includegraphics[width=8cm]{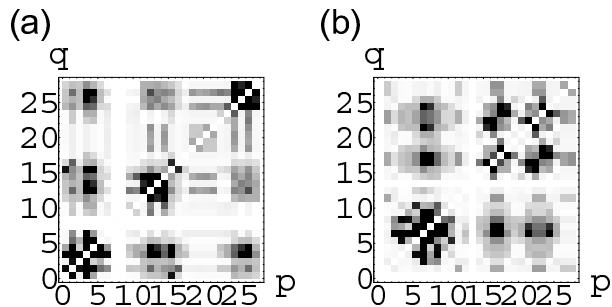}
  \end{center}
\caption{$I_{p,q}$ calculated at $m=600$ for (a) $\rm
F_2$-(\ref{eq:sim-an-param-2})-$D_{2h}$ and (b) $\rm
F_2$-\cite{oers3}-$D_{2h}$ (labels defined in Table
\ref{tab:N2-F2-orderings}).} \label{fig:F2-Ipq}
\end{figure}

For $\rm F_2$, in Fig.~\ref{fig:F2-Ipq} one cannot say that one plot of $I_{p,q}$ is more
compact than the other. Despite this similarity, the differences in
the energy convergence are as pronounced as for CO. Only this time
the criterion of our work leads to the more exact energy.

To conclude, we can say that the application of $I_{p,q}$,
Eqs.~(\ref{eq:simulated-annealing}) and~(\ref{eq:sim-an-param-2})
leads to an ordering with a considerably better energy convergence
than a HF ordering, and the results do not depend on the symmetry used
for the underlying HF calculation. In addition, orderings with a good
energy convergence have the following properties: the bandwidth of
the $I_{p,q}$ matrix is small, large elements $I_{p,q}$ are grouped on
the secondary diagonal, and pronounced accumulation and scattering of
large $I_{p,q}$ elements are avoided.

However, we have not been able to establish a distinct correspondence
between the orbital interaction and the energy convergence, although
$I_{p,q}$ is a very reliable quantity. Therefore, one should test a few different
orderings determined by varying some of the parameters of the
annealing process before one sets up a DMRG
calculation aimed at high accuracy. The resulting energy convergence
can be checked for small sizes of the Hilbert space. This is
demonstrated in Fig.~\ref{fig:energies}, where  the orderings with the
lowest energy for $m=200$ also have the lowest energy for $m=600$. 

\subsection{Change of the basis}

Evidently, it is desirable to obtain additional insight into the
mechanism for energetic convergence in the DMRG. One way to achieve
this is not only to consider the ordering of the orbitals but also the
choice of the orbitals themselves. Since the canonical orbitals are not a
mandatory choice, it is possible to construct a new basis that
might suit the DMRG better. So far, there has been one attempt to use
a localised basis in the DMRG~\cite{White2} that has not, however, led to an
improved convergence relative to canonical orbitals.

An obvious choice are natural orbitals, i.e., the eigenfunctions
of the one-particle density matrix
(Eq.~(\ref{eq:one-particle-denmat})). They lead to rapid
convergence of the configuration interaction scheme, and should
therefore be favourable for the DMRG as well. Since the
one-particle density matrix is contained in $\rho^{pq}$, one can
easily construct approximate natural orbitals, for example, from
an approximate wave function of a calculation at $m=200$ with HF
ordering.

\begin{figure}[ht]
  \begin{center}
    \includegraphics[width= 13cm]{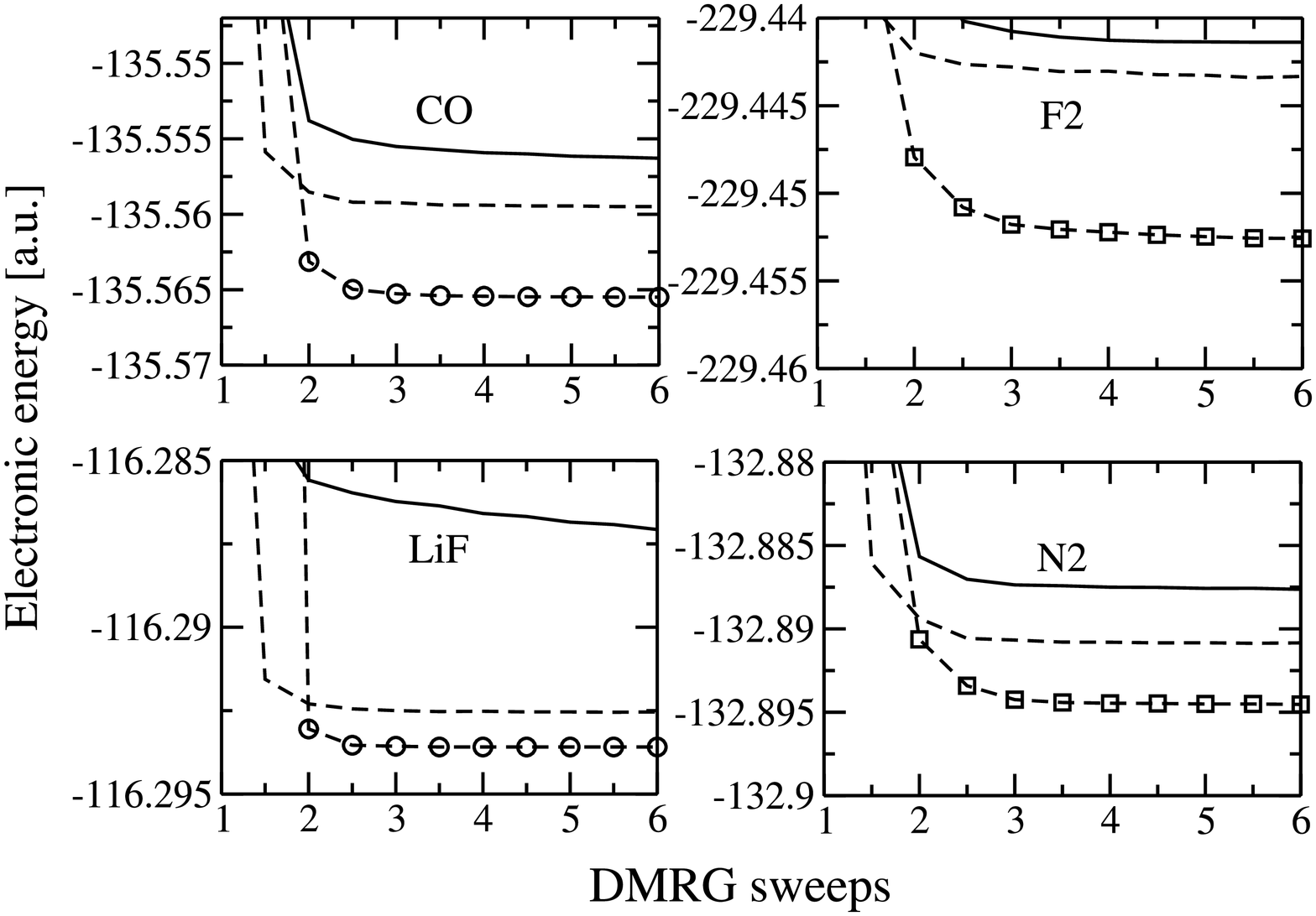}
  \end{center}
\caption{Electronic energies (no nucleus-nucleus interaction)
versus number of DMRG sweeps for $m=200$ (sweeps 1 to 6). Solid
line: canonical orbitals and HF ordering, dashed line: natural
orbitals ordering by eigenvalues, dashed line (circles, squares):
natural orbitals ordering with
Eqs.~(\ref{eq:interaction}),~(\ref{eq:simulated-annealing}),~(\ref{eq:sim-an-param-2})
($C_{2v}$, $D_{2h}$).} \label{fig:nat-200}
\end{figure}

The natural orbitals must also be ordered on the lattice.
One can use an ordering according to occupation
number as a reference ordering.
DMRG results for the electronic energies for the four sample molecules
calculated with canonical and natural orbitals for various orderings are
displayed in Fig.~\ref{fig:nat-200}.
One can see that there is a small increase in convergence for the
natural orbitals compared to the
canonical orbitals in the HF ordering in every case.
One can then apply the ordering criterion of this work
which consistently improves energy convergence. However, the
optimal energy convergence does not exceed that found for
the best orderings of canonical orbitals (not shown in
Fig.~\ref{fig:nat-200}).

Finally, one can consider an iterative improvement of the natural
orbitals: Beginning with a DMRG calculation using canonical
orbitals in HF ordering, one can obtain an improved wave function
which can then be used to calculate natural orbitals which in turn can
serve as a basis for a next DMRG calculation yielding a wave function with an
even lower energy. This procedure can be repeated yielding an
improved wave function and energy at every iteration until the
energy converges. We have applied this iterative procedure and have
found that the energies decrease only for two or three iterations and
then start to fluctuate. Therefore, further investigation is
needed to understand how to construct an optimal basis for the DMRG.

\section{Summary}
\label{sec:summary}

In this work, we have used concepts from quantum information
theory to formulate a definition of orbital interaction $I_{p,q}$.
For a given wave function, $I_{p,q}$ is defined by the
subtraction of the entanglement of two orbitals taken together
with the rest of the system from the sum of the entanglement of
two individual orbitals with the rest of the system. The advantage
of this definition is that one includes information beyond the
Hartree-Fock treatment. The disadvantage is that a correlated wave
function must be calculated. Given a correlated wave function, we
have developed a recipe in Section~\ref{sec:QIT-recipe} for the
calculation of $I_{p,q}$. This recipe additionally provides an
alternative method to calculate one-orbital entropies which are
also of central importance in other work~\cite{oers2,oers3}.

We have calculated $I_{p,q}$ using correlated wave functions obtained
from a DMRG calculation and the recipe of Section~\ref{sec:QIT-recipe}.
The resulting interaction $I_{p,q}$ does not depend strongly on
the accuracy of the underlying wave function. For the four test
molecules we have treated, we have shown that it is possible to
calculate $I_{p,q}$ in a reliable
fashion. The structure of $I_{p,q}$ is consistent with another
criterion based on one-orbital entropies and with chemical
intuition: partially occupied orbitals of the same irreducible
representation interact strongly.

As an application, we have used $I_{p,q}$ to study the ordering
problem in the DMRG, in which one has to order orbitals on a
one-dimensional lattice so that strongly interacting orbitals are
near each other. We have developed a cost function for a simulated
annealing process that leads to an improved ordering for all of
the cases we have treated, i.e., the subsequent DMRG calculation
leads to a lower energy. We have also found that orderings with a
good energy convergence have a small bandwidth in the $I_{p,q}$
matrix, and have a distribution of large matrix elements on the
secondary diagonal that is relatively uniform along the lattice.
However, we have not been able to identify a consistent scheme
that leads to an optimal ordering of the orbitals for the DMRG
with this approach. Careful checks and additional attempts to find
better orderings are needed.

A more general solution to this problem might lie in the
construction of the basis itself. We have therefore investigated
the influence of the use of natural orbitals on the convergence of
the DMRG. This leads to a slight improvement over the use of
canonical orbitals, but the ordering problem still remains. It is
known, however, that the DMRG yields excellent results with a basis of
$p_z$ orbitals for conjugated polymers. Hence, an optimal basis should
consist of orbitals that are localised in real space and are close in
energy. It might then be possible to construct an optimal basis for the DMRG for
which the ordering is either obvious or irrelevant. Then the
ordering problem should be of minor importance. This will be the
topic of a subsequent study.

\ack{J. Rissler acknowledges the support by the Alexander von Humboldt
Foundation.}

\end{document}